\documentclass[aps,prd,twocolumn,showpacs,superscriptaddress,groupedaddress]{revtex4}  % for review and submission
\usepackage{graphicx}  % needed for figures
\usepackage{dcolumn}   % needed for some tables
\usepackage{bm}        % for math
\usepackage{amssymb}
\usepackage{epstopdf}
\usepackage{epsfig}
\usepackage{float}
\usepackage{amsfonts}
\usepackage{amsmath}
\usepackage{hyperref}
\hypersetup{colorlinks,citecolor=blue}

\numberwithin{equation}{section}
\newtheorem{acknowledgement}{Acknowledgement}
\newtheorem{theorem}{Theorem}
\newtheorem{definition}[theorem]{Definition}

\newcommand{\figref}[1]{Fig.~(\ref{#1})}
\newcommand{\reffig}[1]{Fig.~(\ref{#1})}

\newcommand{\figdir}[1]{./Figures/#1}

\begin{document}
\widetext
\leftline{Version 1.0 as of \today}
\leftline{Primary authors: Claudio Durastanti, Yabebal T. Fantaye, Frode K. Hansen, Domenico Marinucci, Isaac Z. Pesenson}
\leftline{To be submitted to Physical Review D}
%\leftline{Comment to {\tt d0-run2eb-nnn@fnal.gov} by xxx, yyy}
\centerline{\em D\O\ INTERNAL DOCUMENT -- NOT FOR PUBLIC DISTRIBUTION}

\title{A Simple Proposal for Radial 3D Needlets}

\author{C. Durastanti} 
\email{durastan@axp.mat.uniroma2.it}
\thanks{supported by the ERC grant \textit{Pascal} 277742}
\affiliation{Dipartimento di Matematica, Università di Roma''Tor Vergata,'' Via della Ricerca Scientifica 1, I-00133 Roma, Italy}
\author{Y. Fantaye} 
\email{fantaye@axp.mat.uniroma2.it}
\thanks{supported by the ERC grant \textit{Pascal} 277742}
\affiliation{Dipartimento di Matematica, Università di Roma''Tor Vergata,'' Via della Ricerca Scientifica 1, I-00133 Roma, Italy}
\author{F. Hansen}
\email{f.k.hansen@astro.uio.no}
\affiliation{Institute of Theoretical Astrophysics, University of Oslo, PO Box 1029 Blindern, 0315 Oslo, Norway }
\author{D. Marinucci}
\email{marinucc@mat.uniroma2.it}
\thanks{partially supported by the ERC grant \textit{Pascal} 277742}
\affiliation{Dipartimento di Matematica, Università di Roma''Tor Vergata,'' Via della Ricerca Scientifica 1, I-00133 Roma, Italy}
\author{I. Z. Pesenson}
\email{pesenson@temple.edu}
\affiliation{Department of Mathematics, Temple University, Philadelphia, PA 19122}

\date{\today}

\begin{abstract}
We present here a simple construction of a wavelet system for the
three-dimensional ball, which we label \emph{Radial 3D Needlets}.
The construction envisages a data collection environment where an
observer located at the centre of the ball is surrounded by
concentric spheres with the same pixelization at different radial
distances, for any given resolution. The system is then obtained
by weighting the projector operator built on the corresponding set
of eigenfunctions, and performing a discretization step which
turns out to be computationally very convenient. The resulting
wavelets can be shown to have very good localization properties in
the real and harmonic domain; their implementation is
computationally very convenient, and they allow for exact
reconstruction as they form a tight frame systems. Our theoretical
results are supported by an extensive numerical analysis.
\end{abstract}

%\newline
%\medskip \textbf{Index Terms:} { } \medskip

\keywords{Data Analysis, Harmonic analysis, wavelets, ball,reconstruction formula, Needlets}

\pacs{98.80.-k, 95.75.Mn, 07.05.Rm}

\maketitle

\section{Introduction}

The last decade has represented a golden era for Cosmology: a
flood of data with unprecedented accuracy has become available on
such diverse fields as Cosmic Microwave Background (WMAP, Planck,
SPT, ACT, see for instance \cite{WMAP, das, hanson, planck23} and
references therein), Ultra High Energy Cosmic Rays (cfr.
\cite{auger}), Gamma rays (Fermi, Agile, ARGO-YBJ+, cfr.
\cite{ARGO, fermi, AGILE}), neutrinos (cfr. \cite{aguilar}) and
many others. Many of these experiments have produced full-sky
surveys, and basically all of them have been characterized by
fields of views covering thousands of square degrees. In such
circumstances, data analysis methods based on flat sky
approximations have become unsatisfactory, and a large amount of
effort has been devoted to the development of procedures which
take fully into account the spherical nature of collected data.

As well-known, Fourier analysis is an extremely powerful method for data
analysis and computation; in a spherical context, Fourier analysis
corresponds to the spherical harmonics dictionary, which is now fully
implemented in very efficient and complete packages such as HealPix, see
\cite{gorski}. For most astrophysical applications, however, standard
Fourier analysis may often result to be inadequate due to the lack of
localization properties in the real domain; because of this, spherical
harmonics cannot handle easily the presence of huge regions of masked data,
nor they can be used to investigate local features such as asymmetries and
anisotropies or the search for point sources.

As a consequence, several methods based on spherical wavelets have
become quite popular in astrophysical data analysis, see for
instance \cite{gonzamart1, donzelli,fay11, Wiaux, cmbneed, pbm,
pietrobon1,rudjord1,starck1, starck2} and also \cite{starcklibro}
for a review. These procedures have been applied to a huge variety
of different problems, including for instance point source
detection in Gamma ray data (cfr. \cite{iuppa1, starckbis}),
testing for nonGaussianity (cfr. \cite{lanm,donzelli,
regan, rudjord1}), searching for asymmetries and local features
(cfr. \cite{cayon, pietrobon1, vielva}), point source
subtraction on CMB data (see \cite{scodeller}), map-making and
component separation (cfr. \cite{basak, delabruille2,
delabrouille1, cardoso}), and several others.

The next decade will probably experience an even more amazing
improvement on observational data. Huge surveys are being planned
or are already at the implementation stage, many of them aimed at
the investigation of the large scale structure of the Universe and
the investigation of Dark Energy and Dark Matter; for instance, a
large international collaboration is fostering the implementation
of the Euclid satellite mission, aimed at a deep analysis of weak
gravitational lensing on nearly half of the celestial sky (see for
instance \cite{EUCLID}). These observational data are also
complemented by N-body simulation efforts (cfr.
\cite{Millennium}) aimed at the generation of realistic
three-dimensional model of the current large scale structure of
the Universe. From the point of view of data analysis, these data
naturally entail a three-dimensional structure, which calls for
suitable techniques of data analysis.

In view of the previous discussion, it is easy to understand the motivation
to develop wavelet systems on the three-dimensional ball, extending those
already available on the sphere. Indeed, some important efforts have already
been spent in this direction, especially in the last few years. Some
attempts outside the astrophysical community have been provided by \cite%
{fmm, michel, pencho}; however the first two proposals are developed in a
continuous setting and do not seem to address discretization issues and the
implementation of an exact reconstruction formula. On the other hand in \cite%
{pencho} the authors proceed by projecting the three-dimensional ball into a
unit sphere in four dimension, and then developing the corresponding
spherical needlet construction in the latter space. While this approach is
mathematically intriguing, to the best of our knowledge it has not led to a
practical implementation, at least in an astrophysical context. This may be
due to some difficulties in handling the required combination of Jacobi
polynomials, and the lack of explicit recipes for cubature points in this
context; moreover the projection of the unit ball on a unit sphere in higher
dimension may induce some local anisotropies, whose effect still needs to be
investigated in an astrophysical context.

Within the astrophysical community, some important proposals for
the construction of three-dimensional wavelets have
been advocated by \cite{starck} and \cite{mcewen}. In the former
paper, the authors propose to use a frequency filter on the
Fourier-Bessel transform of the three-dimensional field. The
proposal by \cite{mcewen} also concentrates on Fourier-Bessel
transforms, and is mainly aimed at the construction of a proper
set of cubature points and weights on the radial part. This is in
practice a rather difficult task: while it is theoretically known
that the cubature points can be taken to be the zeroes of Bessel
functions of increasing degrees, in practice these points are not
available explicitly and the related computations may be quite
challenging. To overcome this issue, in \cite{mcewen} a very
interesting solution is advocated: more precisely, the authors
start by constructing an exact transform on the radial part using
damped Laguerre polynomials, which allow for an exact quadrature
rule. Combining this procedure with the standard spherical
transform, they obtain an exact 3-dimensional decomposition named
Fourier-Laguerre transform. Their final proposal, the so-called
flaglet transform, is then obtained by an explicit projection onto
the Bessel family (e.g., a form of harmonic tiling on the
Fourier-Laguerre transform); this approach is computationally
feasible and exhibits very good accuracy properties from the
numerical point of view.

Our starting point here is to some extent related, and quite
explicitly rooted in the astrophysical applications we have in
mind. In particular, we envisage a situation where an observer
located at the centre of a ball is collecting data, e.g., we
assume that she/he is observing a family of concentric spheres
centred at the origin. At a given resolution level, the
pixelization on each of these spheres is assumed to be the same,
no matter their radial distance from the origin - this seems a
rather realistic representation of astrophysical experiments,
although of course it implies that with respect to Euclidean
distance the sampling is finer for points located closer to the
observer. In this sense, our construction has an implicit radial
symmetry which we exploit quite fully: in particular, we view the
ball of radius $R$ as a manifold $M=[0,R]\times S^{2},$ and we
modify the standard spherical Laplacian so that the distance
between two points on the same spherical shell depends only on the
angular component and not on the radius of the shell. The
corresponding eigenfunctions have very simple expressions in terms
of trigonometric polynomials and spherical harmonics; our system
(which we label \emph{3D radial needlets}) is then built out of
the same procedures as for needlets on the sphere, namely
convolution of a projection operator by means of a smooth window
function $b(\cdot),$ and discretization by means of an explicitly
provided set of cubature points. Concerning the latter, cubature
points and weights arise very simply from the tensor products of
cubature points on the sphere (as provided by HealPix in
\cite{gorski}, for instance), and a uniform discretization on the
radial part, which is enough for exact integration of
trigonometric polynomials.

We believe the present proposals enjoys some important advantages, such as

\begin{enumerate}
\item very good localization properties (in the suitable distance, as
motivated before); these properties can be established in a fully rigorous
mathematical way, exploiting previous results on the construction of
wavelets for general compact manifolds in \cite{gm2}, see also \cite%
{pesenson2};

\item an exact reconstruction formula for band-limited functions, a
consequence of the so-called tight frame property; the latter property has
independent interest, for instance for the estimation of a binned spectral
density by means of needlet coefficients (see \cite{bkmpAoS2} for analogous
results in the spherical case);

\item a computationally simple and effective implementation scheme,
entailing uniform discretization and the exploitation of existing packages;

\item a natural embedding into experimental designs which appear quite
realistic from an astrophysical point of view, as discussed earlier.

\end{enumerate}

The construction and these properties are discussed in more details in
the rest of this paper; we note that the same ideas can be simply
extended to cover the case of spin valued functions, along the same
lines as done for standard 2D needlets by \cite{gm3,gm4}: these
extensions may be of interest to cover forthcoming data on weak
gravitational lensing (e.g. \cite{EUCLID}).

The paper is divided as follows:
Section 2 presents the background material on our embedding of the
three-dimensional ball, related Fourier analysis and
discretization issues; Section 3 presents the 3D radial needlets
construction in details; Section 4 discusses the comparison with
possible alternative proposals; Section 5 presents our numerical
evidence, while some technical computations are collected in the
Appendix.

\section{The Basic Framework}

As mentioned in the Introduction and discussed at length also in
other papers (\cite{mcewenloc, mcewen}), in an astrophysical
framework data collection on the ball is characterized by a marked
asymmetry between the radial part and the spherical component.
Indeed, it is well-known that for astrophysical datasets
observations at a growing radial distance corresponds to events at
higher redshift, which have hence occurred further away in time,
not only in space; data at different redshifts correspond at
different epochs of the Universe and are hence the outcome of
different physical conditions. From the experimental point of
view, the signal-to-noise ratio is strongly influenced by radial
distance; for instance, a strong selection bias is introduced as
higher and higher intrinsic luminosity is needed to observe
objects at growing redshift. The asymmetry between the radial and
spherical components is also reproduced in data storing
mechanisms, which typically adopt independent
discretization/pixelization schemes for the two components.

In view of these considerations, it seems natural and convenient to
represent functions/observations on the three-dimensional ball $\mathcal{B}%
_{R}\mathcal{=}\left\{ (x_{1},x_{2},x_{3}):x_{1}^{2}+x_{2}^{2}+x_{3}^{2}\leq
R\right\} $ as being defined on a family of concentric spheres (shells),
indexed by a continuous radial parameter (i.e., a growing redshift); here,
the radius $R$ of the ball can be taken to represent the highest redshift
value $z$ in the catalogue being analyzed, $R=z_{\max }.$ In the sequel, we
shall work with spherical coordinates $(r,\theta ,\varphi );$ for notational
convenience, we take $r=2\pi z/R,$ so that $r\in \lbrack 0,2\pi ].$
Formally, this means we shall focus on the product space
\begin{equation*}
L^{2}(M,d\mu )=L^{2}((0,2\pi ],dr)\otimes L^{2}(S^{2},\>d\sigma )\text{ ,}
\end{equation*}%
where $d\mu=drd\sigma$, $d\sigma =\left( 4\pi \right) ^{-1}\sin \theta d\phi $ and $dr$
denotes standard Lebesgue measure on the unit interval. This simplifying
step is at the basis of our construction; indeed, for our purposes it will
hence be sufficient to construct a tight and localized frame on $%
L^{2}(M,d\mu ),$ a task which can be easily accomplished as follows.

Recall first that for square-integrable functions on the sphere, e.g. on $%
L^{2}(S^{2},d\sigma ),$ a standard orthonormal basis is provided by the set
of spherical harmonics
\begin{equation*}
\left\{ Y_{\ell ,m}\left( \theta ,\phi \right) \right\} \text{ },\text{ }%
\text{ }\ell =0,1,2,...,m=-\ell ,...,\ell \text{ ,}
\end{equation*}%
where $\theta \in \left[ 0,\pi \right]$  and $\varphi \in \left[ 0,2\pi \right)$.
As well-known, the spherical harmonics provide a complete set of
eigenfunctions for the spherical Laplacian
\begin{equation*}
\Delta _{S^{2}}=\frac{1}{\sin \theta }\frac{\partial }{\partial \theta }%
(\sin \theta \frac{\partial }{\partial \theta })+\frac{1}{\sin ^{2}\theta }%
\frac{\partial }{\partial \varphi },
\end{equation*}
indeed%
\begin{equation*}
\Delta _{S^{2}}Y_{\ell ,m}=-\ell (\ell +1)Y_{\ell ,m}\text{ , }\ell =1,2,....
\end{equation*}%
Hence, for any $f\in L^{2}\left( S^{2},d\sigma \right) $, we have
\begin{equation*}
f\left( \omega \right) =\sum_{\ell \geq 0}\sum_{m=-l}^{l}a_{\ell
,m} Y_{\ell ,m}\left( \omega \right)\text{ , }\omega \in
S^{2}\text{ ,}
\end{equation*}%
where the coefficients $\left\{ a_{\ell m}\right\} $ are evaluated by
\begin{equation*}
a_{\ell ,m}=\int_{S^{2}}\overline{Y}_{\ell ,m}\left( \omega \right) f\left(
\omega \right) \sigma \left( d\omega \right) \text{ .}
\end{equation*}

On the other hand, for the radial part we consider the standard Laplacian
operator $\frac{\partial ^{2}}{\partial r^{2}}$, for which an orthonormal
family of eigenfunctions is well-known to be given, for $n=0,1,2,\ldots$, by%
\begin{equation*}
\frac{\partial ^{2}}{\partial r^{2}}\left( 2\pi \right) ^{-\frac{1}{2}}\exp
\left( inr\right) =-n^{2}\left( 2\pi \right) ^{-\frac{1}{2}}\exp \left(
inr\right) \text{ . }
\end{equation*}%
We can hence define a Laplacian on $M$ by%
\begin{equation*}
\Delta _{M}:=\frac{\partial ^{2}}{\partial r^{2}}+\Delta _{S^{2}}\text{ ,}
\end{equation*}%
e.g.%
\begin{equation}
\Delta _{M}\left( \exp \left( inr\right)
Y_{\ell m}\left( \omega \right) \right) =-e_{n,\ell }\exp \left( inr\right) Y_{\ell m}\left( \omega \right) \text{ , }%
\label{ourlap}
\end{equation}%
where
\begin{equation*}
e_{n,\ell }=\left( n^{2}+\ell (\ell +1)\right) \text{ . }
\end{equation*}%
\ It is interesting to compare the action of $\Delta _{M}$ with the standard
Laplacian in spherical coordinates, which is given by%
\begin{equation}
\Delta =\frac{1}{r^{2}}\frac{\partial }{\partial r}r^{2}\frac{\partial }{%
\partial r}+\frac{1}{r^{2}}\Delta _{S^{2}}\text{ ;}  \label{standlap}
\end{equation}%
it can be checked that $\Delta _{M}$ is the Laplace-Beltrami operator which
correspond to the metric tensor%
\begin{equation*}
g_{M}=\left(
\begin{array}{ccc}
1 & 0 & 0 \\
0 & 1 & 0 \\
0 & 0 & \sin ^{2}\theta
\end{array}%
\right) \text{ ,}
\end{equation*}%
as opposed to the usual Euclidean metric in spherical coordinates%
\begin{equation*}
g=\left(
\begin{array}{ccc}
1 & 0 & 0 \\
0 & r^{2} & 0 \\
0 & 0 & r^{2}\sin ^{2}\theta
\end{array}%
\right) \text{ .}
\end{equation*}%
Likewise, the intrinsic distance between points $x_{1}=\left( r_{1},\omega
_{1}\right) =\left( r_{1},\theta _{1},\varphi _{1}\right) $ and $%
x_{2}=\left( r_{2},\omega _{2}\right) =\left( r_{2},\theta _{2},\varphi
_{2}\right) ,$ $\omega _{1},\omega _{2}\in S^{2},$ $r_{1},r_{2}\in \lbrack
0,2\pi ],$ $x_{1},x_{2}\in M,$ is provided by%
\begin{equation}
d_{M}(x_{1},x_{2})=\sqrt{(r_{1}-r_{2})^{2}+d_{S^{2}}^{2}(\omega _{1},\omega
_{2})}\text{ ,}  \label{Mdistance}
\end{equation}%
as opposed to Euclidean distance in spherical coordinates%
\begin{equation}
d(x_{1},x_{2})=\sqrt{(r_{1}-r_{2})^{2}+r_{1}r_{2}d_{S^{2}}^{2}(\omega
_{1},\omega _{2})}\text{ .}  \label{Edistance}
\end{equation}%
In words, in our setting the distance between two points at a given redshift
is simply equal to their angular separation, whatever the redshift; on the
contrary, under the Euclidean distance for a given angular separation the
actual distance grows with the radial component. It can be argued that the
metric $d_{M}(\cdot ,\cdot )$ is a natural choice for any wavelet
construction where the radial component is decoupled from the spherical one.
Given this choice of metric, our construction can be advocated as optimal,
in the sense that it is based on the eigenfunctions of the associated
Laplacian, and hence can be shown to enjoy excellent localization properties
in the real and harmonic domains.

As a consequence of the previous discussion, the family of functions
\begin{equation}
u_{\ell ,m,n}\left( r,\vartheta ,\phi \right) =\left(2 \pi \right)^{-\frac{1%
}{2}} \exp \left(inr\right)Y_{\ell ,m}\left( \vartheta ,\phi \right)
\label{Mbasis}
\end{equation}%
provides an orthonormal basis on $L^{2}(M,\>d\mu )$, e.g., for any $F\in
L^{2}\left( M,d\mu \right) $, the following expansion holds in $L^{2}(M,d\mu
):$
\begin{equation}
F\left( r,\vartheta ,\phi \right) =\sum_{\ell \geq 0}\sum_{m=-\ell
}^{\ell }\sum_{n\geq 0}a_{\ell ,m,n}u_{\ell ,m,n}\left(
r,\vartheta ,\phi \right) \text{ ,}  \label{usum}
\end{equation}%
where
\begin{eqnarray}
\nonumber a_{\ell ,m,n}&:=&\left\langle F,u_{\ell ,m,n}\right\rangle _{L^{2}(M,d\mu
)}\\
&=&\int_{M}F\left( x\right) \overline{u}_{\ell ,m,n}\left( x\right) d\mu
\left( x\right) \text{ .}  \label{ureconst}
\end{eqnarray}%
Of course, we can also rewrite (\ref{ourlap}) more compactly as
\begin{equation}
\Delta _{M}u_{\ell ,m,n}=-e_{\ell ,n}u_{\ell ,m,n}\text{ .}  \label{D}
\end{equation}

It may be noted that by taking a trigonometric basis for the
radial part, we are implicitly assuming that the functions to
reconstruct satisfy periodic boundary conditions. For
astrophysical applications, this does not seem to bring in any
problem. Indeed, we envisage circumstances where catalogues are
provided within some band of redshift values $0<z_{min}<z_{max}$;
periodicity is then obtained by simply padding zero observations
at the boundaries.

The final step we need to complete our frame construction is
discretization; the procedure is standard, and can be outlined as
follows. Let $\Pi _{\Lambda}$ be the a set of band-limited
functions of order smaller than $\Lambda$, i.e. the linear span of
the basis elements $\left\{ u_{\ell ,m,n}\right\} $ for which the
corresponding eigenvalues are such that $e_{\ell ,n}\leq
\Lambda $. Given an integer $j$, there exists a set of points and weights $%
\aleph _{j}:=\left\{ \xi _{j,q,k}=\left( r_{j,q},\>\theta
_{j,k},\>\varphi _{j,k}\right) \right\} ,$ and positive weights
$\left\{ \lambda _{j,q,k}\right\} ,$ $1\leq q\leq Q_{j}$, $1\leq
k\leq K_{j}$, such that for all $P\in $ $\Pi _{B^{2j+2}}$ the
following exact cubature formula holds:
\begin{eqnarray}
\nonumber \int_{M}P\left(x\right)d\mu\left(x\right) &=&\int_{S^{2}}\int_{0}^{1}P\left( r,\theta ,\varphi \right)
drd\sigma \left( \theta ,\varphi \right)\\
&=&\sum_{q=1}^{Q_{j}}\sum_{k=1}^{K_{j}}P\left( r_{j,q},\>\theta
_{j,k},\>\varphi _{j,k}\right) \lambda _{j,q,k}\text{ ,}  \label{cubature}
\end{eqnarray}%
where the cubature points and weights satisfy
\begin{equation*}
\lambda _{j,q,k}\approx B^{-3j}\text{ , }K_{j}\approx B^{2j}\text{ , }%
Q_{j}\approx B^{j}\text{ ,}
\end{equation*}%
and the notation $x_{1}\approx x_{2}$ means that there exists
$c>0$ such that $c^{-1}x_{1}\leq x_{2}\leq cx_{1}$. More
explicitly, $K_{j}$ denotes the pixel cardinality on the spherical
part, and $Q_{j}$ represents the pixel cardinality on the radial
part for a given resolution level $j$. In words, this means that
for such functions integrals can be evaluated by finite sums over
suitable points without any loss of accuracy. The existence of
cubature points with the required properties follows immediately
by the tensor construction which we described in the previous
subsection: in particular, the spherical component $\left( \theta
_{j,k},\varphi _{j,k}\right) $ can be provided along the same
scheme as in \cite{npw2}, while for practical applications the
highly popular pixelization scheme provided by HealPix (cfr.
\cite{gorski}) may be used; on the radial part, cubature points
maybe simply taken to be given by $r_{j,q}:=\frac{2\pi q}{B^{j}},
q=0,\ldots,\left[B^{j}\right]-1$, where $\left[\cdot\right]$
denotes the integer part, cfr. Section \ref{sec:algorithm}.

\section{3D Radial Needlets and their Main Properties}

Having set the basic framework for Fourier analysis and discretization on $%
L^{2}(M,d\mu ),$ the construction of 3D radial needlets can proceed along
very much the same lines as on the sphere or other manifolds (compare \cite%
{npw1, npw2, pes1, pes2, pg1, pesenson2}). More precisely, let us
fix a scale parameter $B>1$, and let $b\left( u\right) $, $u\in
\mathbb{R}$, be a positive kernel satisfying the following three
assumptions:

\begin{enumerate}
\item $b\left( \cdot \right) $ has compact support in $\left[ 1/B,B\right] $;

\item $b\left( \cdot \right) $ is infinitely differentiable in $\left(
0,\infty \right) $;

\item the following partition of unity property holds%
\begin{equation*}
\sum_{j=-\infty }^{\infty }b^{2}\left( \frac{u}{B^{j}}\right) =1\text{ , for
all }u>0\,\text{\ .}
\end{equation*}
\end{enumerate}

\begin{figure*}[]
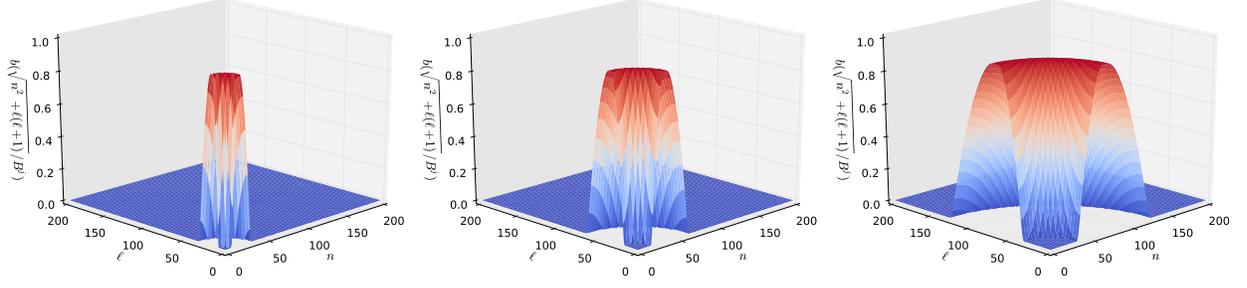

\begin{center}
\includegraphics[width=0.3\textwidth,angle=0]{\figdir{gln_3d_lmax200_nmax200_j4.pdf}} %
\includegraphics[width=0.3\textwidth,angle=0]{\figdir{gln_3d_lmax200_nmax200_j5.pdf}} %
\includegraphics[width=0.3\textwidth,angle=0]{\figdir{gln_3d_lmax200_nmax200_j6.pdf}}
\end{center}
\caption{3D needlet window functions $b\left( \frac{\sqrt{e_{\ell
        ,n}}}{B^{j}}\right)$: $j=4$ (left panel), $j=5$ (middle panel)
  and $j=6$ (right panel). \label{fig:bfunc}}
\end{figure*}

In \reffig{fig:bfunc} we show a visualization of $b\left( \frac{\sqrt{%
e_{\ell ,n}}}{B^{j}}\right)$ for different needlet frequency $j$ values and $ell_{max}=n_{max}=200$.

Numerical recipes for the construction of window functions satisfying the
three conditions above are now well-known to the literature; for instance,
in \cite{pbm} (see also \cite{marpecbook}), the following procedure is
introduced:

\begin{itemize}
\item STEP 1: Construct the $C^{\infty }$-function
\begin{equation*}
\phi _{1}\left( t\right) =\left\{
\begin{array}{c}
\exp \left( -\frac{1}{1-t^{2}}\right) \\
0%
\end{array}%
\right.
\begin{array}{c}
t\in \left[ -1,1\right] \\
\text{otherwise}%
\end{array}%
\text{ ,}
\end{equation*}%
compactly supported in $\left[ -1,1\right] $;

\item STEP 2: Implement the non-decreasing $C^{\infty }$-function
\begin{equation*}
\phi _{2}\left( u\right) =\frac{\int_{-1}^{u}\phi _{1}\left( t\right) dt}{%
\int_{-1}^{1}\phi _{1}\left( t\right) dt}\text{ ,}
\end{equation*}%
normalized in order to have $\phi _{2}\left( -1\right) =0$; $\phi _{2}\left(
1\right) =1$;

\item STEP\ 3: Construct the function
\begin{equation*}
\phi _{3}\left( t\right) =\left\{
\begin{array}{c}
1 \\
\phi _{2}\left( 1-\frac{2B}{B-1}\left( t-\frac{1}{B}\right) \right) \\
0%
\end{array}%
\right.
\begin{array}{c}
t\in \left[ 0,1/B\right] \\
t\in \left( 1/B,1\right] \\
t\in \left( 1,\infty \right)%
\end{array}%
\text{ ;}
\end{equation*}

\item STEP 4: Define, for $u\in\mathbb{R}$%
\begin{equation*}
b^{2}\left( u\right) =\phi _{3}\left( \frac{u}{B}\right) -\phi _{3}\left(
u\right) \text{ .}
\end{equation*}
\end{itemize}

Now recall that $e_{\ell ,n}=n^{2}+\ell (\ell +1)$, and let the symbol $%
\left[ \ell ,n\right] _{j}$ denote the pairs of $\ell $ and $n$ such that $%
e_{\ell ,n}$ is bounded above and below respectively by $B^{2\left(
j+1\right) }$ and $B^{2\left( j-1\right) }$, i.e.
\begin{equation*}
\left[ \ell ,n\right] _{j}=\left\{ l,n:B^{2\left( j-1\right) }\leq e_{\ell
,n}\leq B^{2\left( j+1\right) }\right\} \text{ }.
\end{equation*}%

We have the following

\begin{definition}
The radial 3D-needlets basis is defined by
\begin{eqnarray}
\nonumber \Phi _{j,q,k}\left( r,\vartheta ,\varphi \right)
&=&\sqrt{\lambda _{j,q,k}}%
\sum_{\left[ \ell ,n\right] _{j}}\sum_{m=-\ell }^{\ell }b\left( \frac{\sqrt{%
e_{\ell ,n}}}{B^{j}}\right)\\
&\times& \overline{u}_{\ell ,m,n}\left( \xi
_{j,q,k}\right) u_{\ell ,m,n}\left( x\right) \text{ ,}  \label{3dneed}
\end{eqnarray}%
where $\lambda _{j,q,k}$ and $\xi _{j,q,k}$ denote respectively the pixel
volume and the pixel center.
\end{definition}

Analogously to the related constructions on the sphere or on other
manifolds, radial 3D-needlets can be viewed as the convolution of the
projector operator
\begin{equation*}
Z_{\ell n}(\xi _{j,q,k},x)=\sum_{m}\overline{u}_{\ell ,m,n}\left( x\right)
u_{\ell ,m,n}\left( \xi _{j,q,k}\right)
\end{equation*}%
with the window function $b\left( \cdot \right)$. The properties
of this system are to some extent analogous to the related
construction on the sphere, as illustrated below.

\subsection{The tight frame property}

Let us first recall the notion of tight frame, which is defined to be a
countable set of functions $\left\{ e_{i}\right\} $ such that%
\begin{equation*}
\sum_{i}\beta_i^2(f)=\int_{\mathcal{B}}f\left(
x\right) ^{2}dx \text{ ;}
\end{equation*}%
where the coefficients $\beta_i(f)$ are defined by
\begin{equation}
\beta_i(f) = \int{f(x)e_i(x)dx},
\end{equation}
so that the ``energy'' of the function $f$ is fully conserved in the collection of $\beta_i$'s;  
we refer for instance to \cite{hernandezweiss, pesensongeophysical} and the
references therein for more details and discussions. 
In words, a tight frame can be basically seen as a (possible redundant) basis; indeed we recall
that tight frames enjoy the same reconstruction property as standard
orthonormal systems, e.g.%
\begin{equation*}
f=\sum_{i}\beta_i(f) e_{i}\text{ },
\end{equation*}%
the equality holding in the $L^{2}$ sense.

It is a straightforward consequence of the previous construction that the
set $\left\{ \Phi _{j,q,k}\right\} _{j,k,q}$ describes a tight frame over $%
L^{2}\left( M,d\mu \right) $, so that an exact reconstruction formula holds
in this space; the details of the derivation of this result are collected in
the Appendix. Indeed, let $F\in L^{2}\left( M,d\mu \right) $, i.e. the space
of functions which have finite norm $\Vert .\Vert _{L^{2}(M)}^{2}$, where%
\begin{equation}
\Vert F\Vert _{L^{2}\left( M,d\mu \right) }^{2}:=\int_{0}^{2\pi
}\int_{S^{2}}F^{2}(r,\vartheta ,\varphi )\sin \vartheta drd\vartheta
d\varphi \text{ .}  \label{norm}
\end{equation}%
The 3D-needlet coefficients are defined as
\begin{equation*}
\beta _{j,q,k}:=\beta _{j,q,k}(F)=\int F\Phi _{j,q,k} d\mu,
\end{equation*}%
% \right\rangle
% _{L^{2}\left( M,d\mu \right) }\text{ ,}

or, more explicitly,%
\begin{eqnarray} \nonumber
\beta _{j,q,k}&=&\sqrt{\lambda _{j,q,k}}\sum_{\left[ \ell ,n\right]
_{j}\ }\sum_{m=-\ell }^{\ell }b\left( \frac{\sqrt{e_{\ell
,n}}}{B^{j}}\right)\\
&  \times& a_{\ell ,m,n}u_{\ell ,m,n}\left(\xi_{j,q,k}\right)  \label{coeff} \text{ ,}
\end{eqnarray}%
where $a_{\ell ,m,n}$ is given by (\ref{ureconst}).
The tight frame property then gives%
\begin{equation*}
\Vert F\Vert _{L_{2}\left( M,d\mu \right) }^{2}=\sum_{j\geq
0}\sum_{q=1}^{Q_{j}}\sum_{k=1}^{K_{j}}|\beta _{j,q,k}|^{2};
\end{equation*}%
this property implies also the reconstruction formula
\begin{equation} \label{needRecon}
F\left( x\right) =\sum_{j\geq
0}\sum_{k=1}^{K_{j}}\sum_{q=1}^{Q_{j}}\beta _{j,q,k}\Phi
_{j,q,k}\left( x\right) \text{ ,}
\end{equation}%
see the Appendix for more discussion and some technical
details.

There are some important statistical applications of the tight
frame property. Firstly, the reconstruction property allows the
implementation of denoising and image reconstruction techniques,
for instance on the basis of
the universally known thresholding paradigm (see for instance \cite%
{bkmpAoS2, donoho1, WASA}); in view of the localization properties discussed
in the following paragraph, such denoising techniques will enjoy statistical
optimality properties, in the sense of minimizing the expected value of the
reconstruction error, defined by%
\begin{equation*}
\mathbb{E}\left[ \left\Vert \widehat{F}-F\right\Vert _{L^{2}(M,d\mu )}^{2}\right]
%\end{equation*}
%\begin{equation*}
=\mathbb{E}%
\left[ \int_{M}\left( \widehat{F}(x)-F(x)\right) ^{2}d\mu \left(x\right)
\right]
\end{equation*}%
Here $\mathbb{E}\left[ \cdot \right] $ denotes expected value and $\widehat{F}$ the
reconstructed function in the presence of additive noise with standard
properties. It is important to stress that this reconstruction error is
measured according to the norm introduced in (\ref{norm}), rather than the
usual Euclidean measure in spherical coordinates, where integration is \
performed with respect to the factor $r^{2}\sin \vartheta drd\vartheta
d\varphi .$ In practical terms, this means that the observations at lower
redshift are given a higher weight when performing image denoising; this
appears a rather reasonable strategy, as most astrophysical catalogues are
more complete and less noisy at lower redshift.

We also note that the tight frame property allows an estimator for the
averaged power spectrum of random fields to be constructed by means of the
squared needlet coefficients, along the same lines as for instance \cite%
{bkmpAoS} in the spherical case. More details and further
investigations on all these issues are left for future research.

\subsection{Localization Properties}

It is immediately seen that the functions $\left\{ \Phi _{j,q,k}\left(\cdot
\right) \right\} $ are compactly supported in the harmonic domain; indeed
for any fixed $j,$ as argued before we have that $b(\cdot)$ is non-zero for $%
u\in \left( B^{-1},B\right) $, and hence it follows that $b\left( \frac{%
\sqrt{e_{\ell ,n}}}{B^{j}}\right) \neq 0$ $\ $only for $e_{\ell ,n}\in \left[
B^{2(j-1)},B^{2(j+1)}\right] $. For instance, for $B=\sqrt{2}$ and $j=4$ we
have $8\leq e_{\ell ,n}\leq 32,$ allowing for the pairs
\begin{equation*}
(\ell
,n)=(1,3),(1,4),(1,5),(2,2),(2,3),(2,4),\ldots,(5,1)%
\text{ .}
\end{equation*}%
It is also easy to establish localization in the real domain by
means of general results on localization of needlet-type
constructions for smooth manifolds. In particular, it follows from
Theorem 2.2 in \cite{gm2}, see also \cite{pg1}, that for all $\tau
\in \mathbb{N},$ there exists constants $c_{\tau }$ such
that%
\begin{equation}
\left\vert \Phi _{j,q,k}\left( x\right) \right\vert \leq \frac{c_{\tau }B^{%
\frac{3}{2}j}}{\left( 1+B^{j}d_{M}\left( x,\xi _{j,q,k}\right) \right)
^{\tau }}\text{ ,}  \label{localineq}
\end{equation}%
uniformly over $j,q,k$ and $x.$ It is very important to notice
that the distance at the denominator is provided by equation
(\ref{Mdistance}).

An important consequence of localization can be derived on the
behaviour of the $L^{p}$ norms for the functions $\left\{ \Phi
_{jqk}\left( \cdot\right) \right\} .$ In particular, it can be proved
by standard arguments (as for instance in \cite{npw2}) that, for
all $1\leq p<\infty $,
\begin{equation}
\Vert \Phi _{j,q,k}\Vert _{L^{p}\left( M,d\mu \right) }^{p}:=\int_{M}\left\vert\Phi _{j,q,k}\right\vert^{p}(x)d\mu\left(x\right) \approx B^{\frac{3}{2}\left( p-2\right) j}\text{ , }
\label{Lpbound}
\end{equation}%
while%
\begin{equation*}
\Vert \Phi _{j,q,k}\Vert _{L^{\infty }\left( M,d\mu \right) }\approx B^{%
\frac{3}{2}j}\text{ . }
\end{equation*}%
The result is consistent with the general characterization for the $L^{p}$%
-norm of spherical needlets on $S^{d},$ which is well-known to be given by $%
\Vert \psi _{j,k}\Vert _{L^{p}\left( S^{d}\right) }^{p}\approx B^{\frac{d}{2}%
\left( p-2\right) j};$ here of course $d=3.$ The proof is completely
standard, and hence omitted; we only remark that this characterization of $%
L^{p}$ properties plays a fundamental role when investigating the optimality
of denoising and image reconstruction techniques based on wavelet
thresholding, see again \cite{bkmpAoS2} for further references and
discussion.

\begin{figure*}[]
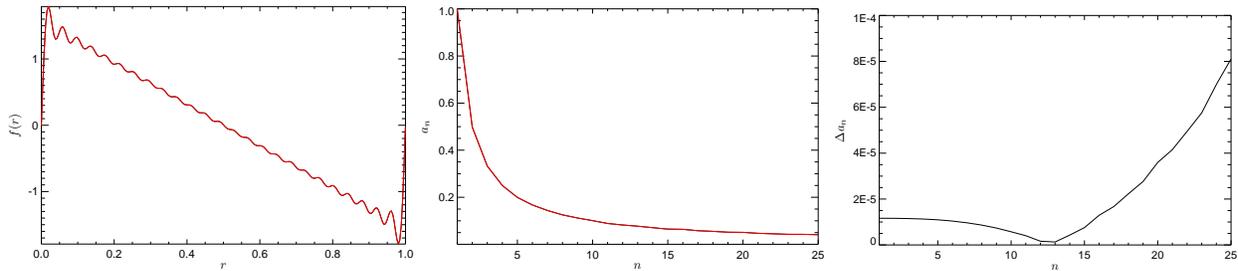

\begin{center}
\includegraphics[width=0.3\textwidth,angle=0]{\figdir{fr_orig_recon.pdf}} %
\includegraphics[width=0.3\textwidth,angle=0]{\figdir{an_orig_recon.pdf}} %
\includegraphics[width=0.3\textwidth,angle=0]{%
\figdir{an_diff_orig_recon.pdf}}
\end{center}
\caption{Radial line synthesis and analysis: input (black curve) and reconstructed (red curve)
function $f(r)$ (left panel), original and reconstructed radial harmonic
coefficient $a_n$ (middle panel), difference between input and reconstructed
$a_n$ (right panel). \label{fig:fr2an}}
\end{figure*}

\section{A comparison with alternative constructions}

The ingredients for the construction of localized tight frames on
a compact manifold are now well-understood; one starts from a
family of eigenfunctions and the associated projection kernel,
then considers a window function to average these projectors over
a bounded subset of frequencies, then proceeds to discretization
by means of a suitable set of cubature points and weights, see for
instance \cite{gm1, gm2,npw1, npw2, pes1, pes2,pesenson2}. The
localization and tight frame properties are then easy consequences
of general results. A natural question then arises, e.g., what
would be the alternative properties of a construction based on a
different choice of eigenfunctions, corresponding for instance to
the standard
Laplacian in spherical coordinates (e.g., (\ref{standlap}) rather than (\ref%
{ourlap})). Indeed, a full system of eigenfunctions for the Laplacian in
spherical coordinates is well-known to be given by
\begin{equation*}
E_{\ell ,m,k}(r,\omega )=\sqrt{\frac{2}{\pi }}\frac{J_{\ell +\frac{1}{2}}(kr)%
}{\sqrt{kr}}Y_{\ell ,m}(\omega )\text{ , }r\in \lbrack 0,1]\text{ , }\omega
\in S^{2},
\end{equation*}%
which can be discretized imposing the boundary conditions $E_{\ell
,m,k}(1,\omega )\equiv 0,$ yielding the family $\left\{ E_{\ell ,m,k_{\ell
p}}(r,\omega )\right\} ,$ for $k_{\ell p}$ the zeroes of the Bessel function
$J_{\ell +\frac{1}{2}}(\cdot)$ in the interval $(0,1).$ 
Writing $e(\ell,k_{\ell_p})$ for the corresponding set of eigenvalues, a needlet type construction would then lead to
the following proposal:%
\begin{eqnarray*}
\Psi _{j,q,k}(r,\omega )&=&\sqrt{\lambda _{j,q,k}}\sum_{\ell ,m}\sum_{k_{\ell_p}}b(%
\frac{\sqrt{e(\ell,k_{\ell_p})}}{B^{j}})\\
&\times& E_{\ell,m,k_{\ell_p}} (r,\omega
)E_{\ell,m,k_{\ell_p}} (r_{q},\omega _{j,k})\text{ ,}
\end{eqnarray*}%
which is close to the starting point of the construction in \cite{mcewen},
the main difference being that their weight function is actually a product
of a radial and spherical part, and its argument is not immediately related
to the eigenvalues of the summed eigenfunctions (see also \cite{starck} for 3D isotropic wavelets based on Bessel functions). It is then easy to show
that $\Psi _{j,q,k}(\cdot,\cdot)$ enjoys a related form of localization
property, namely for all $\tau \in \mathbb{N},$ there exists constants $%
c_{\tau }$ such that%
\begin{equation}
\left\vert \Psi _{j,q,k}(x)\right\vert \leq \frac{c_{\tau }B^{\frac{3}{2}j}}{%
\left( 1+B^{j}d\left( x,\xi _{j,q,k}\right) \right) ^{\tau }}\text{ , }%
\label{loc3d}
\end{equation}%
where $x=(r,\omega)$, $\xi _{j,q,k}=(r_{q},\xi _{jk})$ and $d(\cdot,\cdot)$ denotes as before standard Euclidean distance. It is
then important to stress the different merits of this construction, with
respect to the one we focussed on earlier:

1) for the system $\left\{ \Psi _{j,q,k}\right\} $ distances are evaluated
with a standard Euclidean metric, and the centre of the ball represents a
mere choice of coordinates, e.g., the radial part depends just on the choice
of coordinated and does not necessarily correspond to a specific physical
meaning

2) for the system $\left\{ \Phi _{j,q,k}\right\} $ distances are very much
determined by the choice of the centre and the radial part does have a
specific intepretation, e.g., the distance to the observer.

The rationale underlying 2) was already explained earlier in this paper; we
envisage a situation where an observer located at the centre of a ball
observes a surrounding Universe made up of concentric spheres having the
same pixelizations. As a consequence "closer" spheres, e.g. those
corresponding to smaller radii, are observed at the same angular resolution
as more distant ones - in Euclidean coordinates, this obviously means that
the sampling is finer. Our construction of radial needlets is simply
reflecting this basic feature: as a consequence, for any given frequency
needlets are more localized in proper distance for points located in spheres
closer to the observer. This may appear quite rational, to the extent in
which for these closer regions the signal-to-noise ratio is higher and hence
the reconstruction may proceed to finer details than in the outer regions.
The practical performance of these ideas is tested in the Section to follow.

\section{Numerical Evidence\label{sec:algorithm}}

In this Section we will describe the numerical implementation of
the algorithm developed in this paper. As far as the spherical
component is concerned, our code exploits the well-tested HEALpix~
pixelization code (cfr. \cite{gorski}), while for the radial part
we use an equidistant pixelization, which is extensively tested
below. To investigate the numerical stability and precision
properties of the construction we propose, we shall analyze an
input band-limited function on the ball which we simulate using a
radial and angular test power spectrum.

\begin{figure*}[]
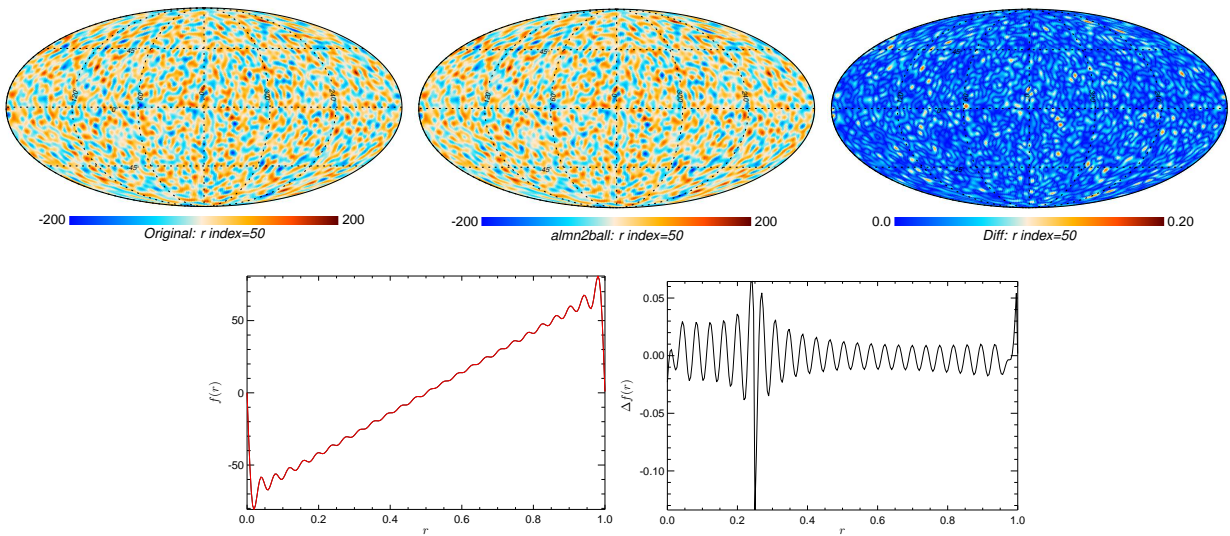

\begin{center}
\includegraphics[width=0.30\textwidth,angle=0]{\figdir{ball_r51.pdf}} %
\includegraphics[width=0.30\textwidth,angle=0]{\figdir{almn2ball_r51.pdf}}
\bigskip
\includegraphics[width=0.30\textwidth,angle=0]{\figdir{ball_almn2ball_diff_r51.pdf}} \\ 
\includegraphics[width=0.30\textwidth,angle=0]{\figdir{fr_ball_almn2ball_ipix201.pdf}}
\includegraphics[width=0.30\textwidth,angle=0]{\figdir{fr_ball_almn2ball_diff_ipix201.pdf}}
\end{center}
\caption{Harmonic space synthesis and analysis: input function at the 51th
shell (top left panel), reconstructed function at the 51th shell (top middle
panel), difference between input and reconstructed functions at the 51th
shell (top right panel), input and reconstructed radial function at the
201 HEALpix~ pixel (bottom left panel), and difference between input and
reconstructed radial function at the 201 HEALpix~ pixel (bottom right panel). \label{fig:ball2almn}}
\end{figure*}

\subsection{Radial reconstruction}

Given a function $g \in L^2\left(\left[0,2 \pi\right],dr \right)$,
we obtain its radial harmonic coefficients by decomposing it in
terms of radial eigenfunctions, i.e. for $n=1,2,\dots$

\begin{equation}
a_{n}=\int_{0}^{2 \pi}{f(r)\frac{\exp (-inr)}{\sqrt{2\pi }}dr}\text{ ,}
\label{eqn:fr2an}
\end{equation}

The reconstruction of the input function can then be obtained via
\begin{equation}  \label{eqn:an2fr}
f(r) = \sum_{n=0}^{ \infty }{a_n(f)\frac{exp(inr)}{\sqrt 2\pi}} \text{.}
\end{equation}

%\left( r\right)

By the standard Euler identity $\exp (inr)=\cos nr+i\sin nr,$ the integral in %
\eqref{eqn:fr2an} involves just sines and cosines function. The
Filon-Simpson algorithm developed by \cite{rosenfeld} can solve
such integral with arbitrary precision for an increasing radial
grid.

To test our radial integration, we used as input the $a_n$ coefficients
shown in black curves in \figref{fig:fr2an} and the reconstruction formula %
\eqref{eqn:an2fr} to obtain $f(r)$ which we evaluate at $N_r=256$ points. After
performing the forward and backward transformation, the
reconstructed $g$ and $a_n$ as well as their differences with
respect to the input values are shown in red color curves in \figref{fig:fr2an}.
The differences in the radial spectra are smaller than $1e-4$ at
all radial multipole values. This validates our radial analysis
and synthesis routines.

\subsection{Discretization of functions on the unit ball}

To describe the reconstruction of a band-limited function through the
pixelization here introduced, we start from an angular power spectrum
$C_{\ell}$ which is derived from CAMB \cite{lewis} $\Lambda$CDM 3D
matter power spectrum at redshift $z=0.5$ and projected to 2D through
Limber approximation; using Healpix, we then generated a set of random
spherical harmonic coefficients from this power spectrum. We
arbitrarily fixed the maximum angular multipole to $\ell_{%
  \mathrm{max}}=65$. For the radial component, we chose
$n_{\mathrm{max}}=25$ and generated a set of Fourier coefficients
$a_n$ which are plotted in the middle box of \figref{fig:fr2an}. We
then convolved the corresponding radial and spherical components to
obtain the desired function on the 3D ball. Our test function $f(r)$
is evaluated at $N_r=256$ radial points while the angular part is defined on HEALpix~ $N_{\mathrm{side}}=64$ pixels.

%The plot of this power spectrum is shown in \reffig{fig:cambclm}

Our numerical 3D grid has a total of $N_r*12*N_{\mathrm{side}}^2$
pixels. We stress again that our main interest here is to
test the accuracy of the codes, so at this stage we are not
concerned with the physical interest of the functions to be
reconstructed; the analysis of more physically motivated models,
such as for instance maps from $N-$body simulations will be
reported in another paper.

% \begin{figure*}[]
% \begin{center}
% \includegraphics[width=0.30\textwidth,angle=0]{\figdir{ball_r51.pdf}} 
% \includegraphics[width=0.30\textwidth,angle=0]{\figdir{beta2ball_r51.pdf}}
% \includegraphics[width=0.30\textwidth,angle=0]{\figdir{ball_beta2ball_diff_r51.pdf}} \\
% \bigskip
% \includegraphics[width=0.30\textwidth,angle=0]{\figdir{fr_ball_beta2ball_ipix201.pdf}}
% \includegraphics[width=0.30\textwidth,angle=0]{\figdir{fr_ball_beta2ball_diff_ipix201.pdf}}
% \end{center}
% \caption{3D needlet space synthesis and analysis: input function at the 51th
% shell (top left panel), reconstructed function at the 51th shell (top middle
% panel), difference between input and reconstructed functions at the 51th
% shell (top right panel), input (black curve) and reconstructed (red curve) radial function at the
% 201 HEALpix~ pixel (bottom left panel), and difference between input and
% reconstructed radial function at the 201th HEALpix~ pixel (bottom right panel). \label{fig:ball2beta}}
% \end{figure*}

\begin{figure*}[]
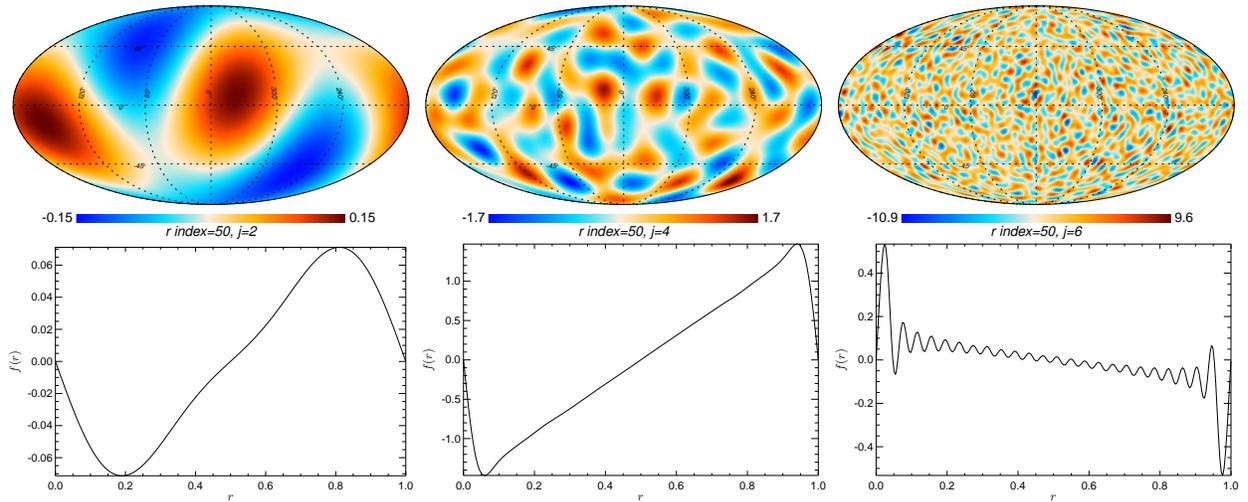

\begin{center}
\includegraphics[width=0.3\textwidth,angle=0]{%
\figdir{ball2beta_r51_jmap2.pdf}} \includegraphics[width=0.3%
\textwidth,angle=0]{\figdir{ball2beta_r51_jmap4.pdf}} %
\includegraphics[width=0.3\textwidth,angle=0]{%
\figdir{ball2beta_r51_jmap6.pdf}}\\[0pt]
\includegraphics[width=0.3\textwidth,angle=0]{%
\figdir{fr_ball2beta_ipix201_jmap2.pdf}} \includegraphics[width=0.3%
\textwidth,angle=0]{\figdir{fr_ball2beta_ipix201_jmap4.pdf}} %
\includegraphics[width=0.3\textwidth,angle=0]{%
\figdir{fr_ball2beta_ipix201_jmap6.pdf}}
\end{center}
\caption{3D needlet component maps: the upper panel shows angular part of
the corresponding needlet component map at the 51th shell while the lower
panel is for the radial part at the 201 HEALpix~ pixel. The $j^{\mathrm{th}}$
component have a compact support of multipoles $B^{j-1}<\protect\sqrt{\ell
(\ell +1)+n^{2}}<B^{j+1}.$ \label{fig:jmaps}}
\end{figure*}

In a similar spirit to the highly popular spherical algorithms
proposed by \cite{gorski}, we have developed the ball equivalent of
the HEALpix~ \emph{alm2map} and \emph{map2alm} codes, which are named
\emph{almn2ball} and \emph{ball2almn}. These two routines, respectively,
solve the analysis and synthesis equations of \eqref{ureconst} and
\eqref{usum}. Equivalently, we have \emph{ball2beta} and \emph{beta2ball} to solve equations \eqref{coeff} and
\eqref{needRecon}, which are respectively performing needlet space
analysis and synthesis.  We have optimized these codes so that they
are fast and run either in serial or parallel mode; we have fully
exploited the rigorously tested and well known HEALpix~ routines, so
that researchers familiar with HEALpix~ should find our code rather
intuitive. The modularity of our code, moreover, means that adapting
our routines to any other pixelization packages will be
straightforward.

To test the accuracy of our code, both in harmonic and real space, in 
\figref{fig:ball2almn} we show the original and reconstructed function
on the unit ball sliced at the 50th radial shell and the radial
function at the 200th HEALpix~ pixel. It can be checked from the
real-space difference plots in the same figure that the residuals
are very small and comparable to HEALpix~'s accuracy. The
difference in harmonic space is such that both the radial and
angular spectra are recovered with accuracy $<1e-4$ at all
multipoles.

\subsection{Radial 3D needlet synthesis and analysis}

Here we start from the $a_{\ell ,m,n}$ coefficients obtained from the
previous section analysis and we compute the needlet coefficients $\beta
_{j,q,k}$ through our routine \emph{almn2beta}, which implements %
\eqref{3dneed}. Optionally one can start directly from the discretization of
the function on the ball and call $ball2beta$ to get $\beta _{j,q,k}$
directly. The needlet parameters we use in this analysis are $j=0,1,..N_{j}-1
$, $q=0,1,..,Q_{j}$, $n_{k}=0,1,..,K_{j}-1$ where ,$N_{j}=7$, $%
Q_{j}=N_{r}=256$, and $K_{j}=12\ast 64^{2}$. To reconstruct the function on
the ball from the needlet coefficients we call $beta2ball$. 
% The comparison
% of the input and reconstructed functions from this analysis is shown in 
% \figref{fig:ball2beta}. 
The pixel by pixel and harmonic space accuracy
of the function reconstructed from needlet coefficients are almost
identical to that of the function reconstructed just from the
$a_{\ell ,m,n}$ coefficients, thus providing some very reassuring
evidence on the accuracy of the algorithm. As mentioned earlier,
further evidence on simulations with more realistic experimental
conditions is left to future research.

In \figref{fig:jmaps}, we show the needlet component maps. The $j^{%
\mathrm{th}}$ needlet component map have a compact support on the range of
combined radial and angular multipoles such that $B^{j-1}<\sqrt{%
\ell(\ell+1)+n^2}<B^{j+1}$.

\appendix

\section{Some technical details}

In this Appendix, we discuss the tight frame property which characterizes
this construction. Indeed, let $F\in L^{2}\left( M,d\mu \right) $: the
corresponding 3D-needlet coefficients are given by (\ref{coeff}). For all $%
1\leq q\leq Q_{j},1\leq k\leq K_{j},$ we have
\begin{eqnarray*}
&&|\beta _{j,q,k}|^{2}\\
&=&\lambda _{j,q,k}\left\vert \sum_{\left[ \ell ,n\right]
_{j}}\sum_{|m|\leq \ell }b\left( \frac{\sqrt{e_{\ell ,n}}}{B^{j}}\right)
a_{\ell ,m,n}u_{\ell ,m,n}\left( \xi _{j,q,k}\right) \right\vert ^{2}\text{ .%
}
\end{eqnarray*}%
From the spectral theorem (see for instance \cite{davies}, cfr. also \cite%
{pesenson2}) we have also, for $F\in L^{2}(M,d\mu )$
\begin{equation*}
\sum_{j\in \mathbb{Z}}\Vert b(B^{-j}\sqrt{\Delta _{M}})F\Vert _{L^{2}\left(
M,d\mu \right) }^{2}=\Vert F\Vert _{L^{2}\left( M,d\mu \right) }^{2}\text{ ,}
\end{equation*}%
where $\Vert .\Vert _{L^{2}(M)}^{2}$ denotes as usual the $L^{2}$-norm of
the function which we recalled in (\ref{norm}). Let us introduce also the
kernel%
\begin{equation*}
\mathcal{K}_{j}\left( x,x^{\prime }\right) :=\sum_{\left[ l,n\right]
_{j}}\sum_{|m=-l}^{l}b\left( \frac{\sqrt{e_{n,l}}}{B^{j}}\right) \overline{u}%
_{\ell ,m,n}\left( x^{\prime }\right) u_{\ell ,m,n}\left( x\right) \text{ , }%
\end{equation*}%
$x,x^{\prime }\in M$, from which we obtain the projections
\begin{eqnarray*}
F_{j}(x) &:=&b(B^{-j}\sqrt{\Delta _{M}})F=\left\langle \mathcal{K}_{j}\left(
.,x\right) ,F(\cdot)\right\rangle _{L^{2}\left( M,d\mu \right) } \\
&=&\sum_{\left[ \ell ,n\right] _{j}}\sum_{|m=-\ell }^{\ell }b\left( \frac{%
\sqrt{e_{\ell ,n}}}{B^{j}}\right) a_{\ell ,m,n}u_{\ell ,m,n}\left( x\right)
\text{  .}
\end{eqnarray*}%
Next,
\begin{eqnarray*}
\nonumber&& \Vert F_{j}(x)\Vert _{L^{2}\left( M,d\mu \right) }^{2}\\
&=&\int_{M}\left\vert
\sum_{\left[ \ell ,n\right] _{j}}\sum_{|m=-\ell }^{\ell } b\left( \frac{\sqrt{%
e_{\ell ,n}}}{B^{j}}\right) a_{\ell ,m,n}u_{\ell ,m,n}\left( x\right)
\right\vert ^{2}d\mu \left( x\right) \text{ .}
\end{eqnarray*}%
The integrand function clearly belongs to $\Pi _{B^{2\left( j+1\right) }},$
i.e. the space of functions which can be expressed as linear combinations of
basis elements corresponding to eigenvalues smaller than $B^{2\left(
j+1\right) }.$ For these functions, an exact cubature formula holds, and we
get
\begin{eqnarray*}
\Vert F_{j}(x)\Vert _{L^{2}\left( M,d\mu \right)
}^{2}&=&\sum_{q=1}^{Q_{j}}\sum_{k=1}^{K_{j}}\lambda _{j,q,k} \left\vert  F_{j}(\xi_{j,q,k}) \right\vert^2 \text{ .}
\end{eqnarray*}%
Therefore we have that%
\begin{eqnarray*}
\Vert F\Vert _{L_{2}\left( M,d\mu \right) }^{2} &=&\sum_{j\geq 0}\Vert
b(B^{-j}\sqrt{\Delta _{M}})F\Vert _{L^{2}\left( M,d\mu \right)
}^{2}\\&=&\sum_{j\geq 0}\Vert F_{j}(x)\Vert _{L^{2}\left( M,d\mu \right) }^{2} \\
&=&\sum_{j\geq 0}\sum_{q=1}^{Q_{j}}\sum_{k=1}^{K_{j}}|\beta _{j,q,k}|^{2},
\end{eqnarray*}%
as claimed.

\begin{acknowledgement}
The authors thank R. Rosenfelder for sending us his Fortran 90 Filon-Simpson code.
\end{acknowledgement}

\end{document}